\let\accentvec\vec
\documentclass[epj]{svjour}

\let\vec\accentvec
\usepackage{graphicx,color,amsmath, amssymb,cite,multirow}
\usepackage[english]{babel}
\usepackage{t1enc}
\usepackage[latin2]{inputenc}
\begin{document}

\title{Tagging of $\eta'$ decay products to analyze chiral restoration}

\author{Máté Csanád\inst{1} \and Mónika Kőfaragó\inst{1}}
\institute{Eötvös University, Department of Atomic Physics, Pázmány Péter s. 1/A, H-1117 Budapest, Hungary}

\date{\today}
\abstract{In case of chiral U$_A(1)$
symmetry restoration the mass of the $\eta'$ boson (the ninth, would-be
Goldstone boson) is decreased, thus its production cross section is
heavily enhanced. The $\eta'$ decays (through one of its decay channels)
into five pions. These pions will not contribute to Bose-Einsten correlations,
thus the production enhancement changes the strength of two-pion
correlation functions at low momentum. Preliminary
results on Bose-Einstein correlation functions support the mass decrease of the $\eta'$ 
boson. In this paper we propose a method to select pions originating from $\eta'$ decays.
We investigate the efficiency of the proposed kinematical cut in several
collision systems and energies with several simulators.
We prove that our method can be used in all investigeted collision systems.}

\maketitle
\section{Introduction}

In relativistic gold-gold collisions of the Relativistic Heavy Ion Collider (RHIC) a strongly interacting quark gluon plasma is produced~\cite{Adcox:2004mh}.
The temperature of this matter may reach values up to 300-600 MeV~\cite{Adare:2008fqa}. At these very high temperatures the degrees of freedom are not hadrons but
quarks or gluons. It is expected that the broken symmetries of QCD may be partially restored in this matter.

In a three-quark QCD there is a $U_L(3)\times{U}_R(3)$ chiral symmetry. According to group theory, $U(3)=SU(3)\times U(1)$, thus chiral
symmetry can be written as $SU_L(3)\times{SU}_R(3)\times{U}_A(1)\times{U}_V(1)$. From this $SU_L(3)\times{SU}_R(3)$ is
the flavor-symmetry, which is spontaneously broken to $SU_V(3)$. With this symmetry breaking eight low-mass Goldstone bosons are created, associated with
the eight pseudoscalar mesons: three pions, three kaons and the $\eta$. The $U_A(1)$ part is also broken, the ninth light Goldstone-boson is
however missing~\cite{Kapusta:1995ww,Csorgo:2009pa}.
This puzzle is solved by the Adler-Bell-Jackiw anomaly: the $U_A(1)$ part of chiral symmetry is broken explicitly due to instantons
tunneling between topologically different QCD vacuum states. The Goldstone-boson appearing with this symmetry breaking
is expected to be massive. It is associated with the $\eta'$ meson, which has a mass of 958 MeV, significantly
higher than that of the other 8 pseudoscalar mesons.

However, in case of chiral symmetry restoration, if the $U_A(1)$ symmetry is partially restored, the mass of the $\eta'$ is
decreased~\cite{Kapusta:1995ww,Csorgo:2009pa}. It is an important aspect of this picture that the symmetry has to be still
partially restored when the $\eta'$ mesons are created, i.e.\ the quark-hadron transition has to happen earlier than the chiral transition. This seems
to be supported by lattice QCD calculations~\cite{Fodor:2009ax}.

\section{Chiral symmetry restoration and mass modification of the $\eta'$}
As mentioned above, chiral symmetry may be partially restored in hot QCD matter, and the mass of $\eta'$ meson might be lower than it's original mass, 958 MeV.
However, the production cross sections of the light mesons are exponentially suppressed by their mass. Hence without mass modification roughly two orders of
magnitute less $\eta'$ mesons are produced than pions. In contrast, decreased mass $\eta'$ mesons will be created more abundantly. The enhancement of $\eta'$
production, according to the Hagedorn formula, may be given as~\cite{Csorgo:2009pa}:
\begin{align}
f_{\eta'}=\left(\frac{m^*_{\eta'}}{m_{\eta'}}\right)^{1-d/2}e^{-\frac{m_{\eta'}-m^*_{\eta'}}{T_{cond}}}
\end{align}
if the mass of the $\eta'$ $m_{\eta'}$ is decreased to $m^*_{\eta'}$. Here $T_{cond}$ is the temperature of the medium when the $\eta'$ mesons are created,
while $d$ is the effective dimension of the expansion. Thus the number of $\eta'$ mesons is closely related to their mass.

The thermalized quark gluon plasma is created at RHIC in gold-gold collisions~\cite{Adcox:2004mh} roughly 1 fm/$c$ after the collision. The matter expands for a time
estimated to be around 6-10 fm/$c$ and cools down to the quark-hadron transition temperature range 150-170 MeV~\cite{Fodor:2009ax}. The hadrons created at this
point are then propagating freely to the detectors. Some of them however decay throughout their journey, as is the $\eta'$ meson. It has a mean lifetime of 1000~fm/$c$,
thus during its life the medium dissolves and the symmetry is broken again, so the $\eta'$ mass increases again at the expense of its momentum:
\begin{align}
{m^*_{\eta'}}^2 +{p^*_{\eta'}}^2={m_{\eta'}}^2 +{p_{\eta'}}^2 \label{e:massshell}
\end{align}
where the starred quantities refer to the in-medium properties, while the others to the vacuum properties of the $\eta'$. Thus the vacuum momentum of the
$\eta'$ will be significantly lower than its original momentum.

The decay of the $\eta'$ happens after it regained its vacuum mass. One important decay channel is the decay into two leptons, $\eta'\rightarrow{l^++l^-}$,
this is investigated in ref.~\cite{Adare:2009qk}. It turns out that there is an excess in the dilepton spectrum at low invariant mass, and this excess
might be related to the $\eta'$ enhancement. There is also a decay mode when the $\eta'$ goes into an $\eta$ and two pions, and the $\eta$ also decays
into three pions:
\begin{align}
\eta'\rightarrow\eta+\pi^++\pi^-\rightarrow\left(\pi^++\pi^-+\pi^0\right)+\pi^++\pi^-\label{e:decay}
\end{align}
and the overall probability of this decay chain is 10\%~\cite{Nakamura:2010zzi}. The average momentum of the resulting five pions is $138$ MeV
due to the low momentum of the original $\eta'$~\cite{Vance:1998wd}. Thus the $\eta'$ decays will have an effect on pion correlation functions.
We will investigate this in the next section.

\section{Two-pion Bose-Einstein correlations}

Final state effects distort two-particle correlation functions. One of the most important final state effect is that of Bose-Einstein correlations.
These can be reviewed as follows. Definition of the two-particle correlation function is:
\begin{align}
C_2\left(p_1,p_2\right)=\frac{N_2\left(p_1,p_2\right)}{N_1\left(p_1\right){N_1\left(p_2\right)}}
\end{align}
where $p_1$ and $p_2$ are the momenta of the two particles, $N_1$ and $N_2$ are the one- and two-particle invariant momentum distributions.
They are defined as:
\begin{align}
N_1(p)&=\int S(x,p) |\Psi_1|^2 d^4x\\
N_2(p_1,p_2)&=\int S(x_1,p_1)S(x_2,p_2) |\Psi_{1,2}|^2 d^4x_2d^4x_1
\end{align}
with $S\left({x,p}\right)$ being the hadronic source function (sometimes noted as emission function), $\Psi_1(x,p)$ is the one-particle wave function
and $\Psi_{1,2}(x_1,x_2,p_1,p_2)$ is the two-particle wave function. Latter has to be symmetrized in case of identical boson pairs. In case of the plain-wave
approximation (neglecting all other final state interactions), we arrive at the following result (see details in ref.~\cite{Csorgo:1999sj}):
\begin{align}
C_2(p_1,p_2)= 1+\frac{\widetilde{S}(q,p_1){\widetilde{S}(q,p_2)}^*}{\widetilde{S}(0,p_1){\widetilde{S}(0,p_2)}^*}
\end{align}  
where $q=p_1-p_2$ and $\widetilde{S}(q,p)$ is the Fourier transformed of the source function (the Fourier transformation is in $x\rightarrow q$).
Introducing $K=(p_1+p_2)/2$ and taking into account $p_1\simeq p_2$, we get:
\begin{align}
C_2(q,K)=1+\frac{{\vert\widetilde{S}(q,K)\vert}^2}{{\vert\widetilde{S}(0,K)\vert}^2}.\label{eq:korrelacio}
\end{align}  

In the core-halo model~\cite{Csorgo:1999sj}, the hadronic source is divided into two parts: a core and a halo. The core consists of the
primordial particles and decay products of very short lifetime resonances. This part of the source has a small size: roughly 10 fm at
maximum. The halo consists then of decay products of long lived resonances, such as $\eta$, $\eta'$ or $K^0_S$. The halo hadrons
are created very far from the core (note the very large lifetime of the previously mentioned particles). When measuring correlation
functions however, due to finite momentum resolution of the detectors very small momentum differences cannot be resolved, i.e.
pairs with such similar momenta are regarded as one by the detectors. In the Fourier transformation, large sizes correspond to
small momenta, thus the halo correlations are not seen in measurements. Therefore we introduce the core source function,
and denote it by $S_C(q,K)$ (and use the $H$ subscript for the halo part). It holds that
\begin{align}
\widetilde{S}(q,K) = \widetilde{S}_C(q,K) + \widetilde{S}_H(q,K)
\end{align}
However, for measureable $q$ values (at least several MeV values), $\widetilde{S}(q,K) = \widetilde{S}_C(q,K)$
(again, large size of $S_H$ corresponds to small width of $\widetilde{S}_H$). Let then the number of particles in the core
be $N_C$, number of particles in the halo be $N_H$. Clearly, for $q=0$:
\begin{align}
\widetilde{S}(0,K) = \frac{N_C+N_H}{N_C}\widetilde{S}_C(0,K)
\end{align}
thus finally 
\begin{align}
C_2(q,K)=1+\lambda\frac{{\vert\widetilde{S}_C(q,K)\vert}^2}{{\vert\widetilde{S}_C(0,K)\vert}^2}
\end{align}
with $\lambda$ being
\begin{align}
\sqrt{\lambda}=\frac{N_C}{N_C+N_H}\label{eq:lambda_def}
\end{align}
Thus
\begin{align}
C_2(q\rightarrow 0,K)={1+\lambda_*}
\end{align}
The $\lambda$ parameter is thus the $q\rightarrow 0$ extrapolated value of the correlation function $C_2$, and it depends on the ratio of the core to the halo.
Thus if the mass of the $\eta'$ is decreased, more of it are produced (see in the previous section), their decay pions will be also be enhanced in number,
so the halo will be larger. This means, that the $\lambda$ parameter is decreased~\cite{Vance:1998wd}. Hence $\eta'$ mass and $\lambda$ value are connected.

It was found~\cite{Csanad:2005nr,Csorgo:2009pa,Vertesi:2009ca} that the $\lambda$ parameter is indeed
decreasing at the kinematical domain of $\eta'$ decay pions. However, it is not experimentally proven that
the $\eta'$ decay pions are causing the decrease. In this paper we investigate a method to kinematically filter out pions from
$\eta'$ decays. If applied to the experimental sample, in case of an $\eta'$ mass modification the $\lambda$ decrese will vanish.

\section{Kinematical domain of pions from $\eta'$ decays}

\begin{figure}
\begin{center}
\includegraphics[width=0.45\linewidth]{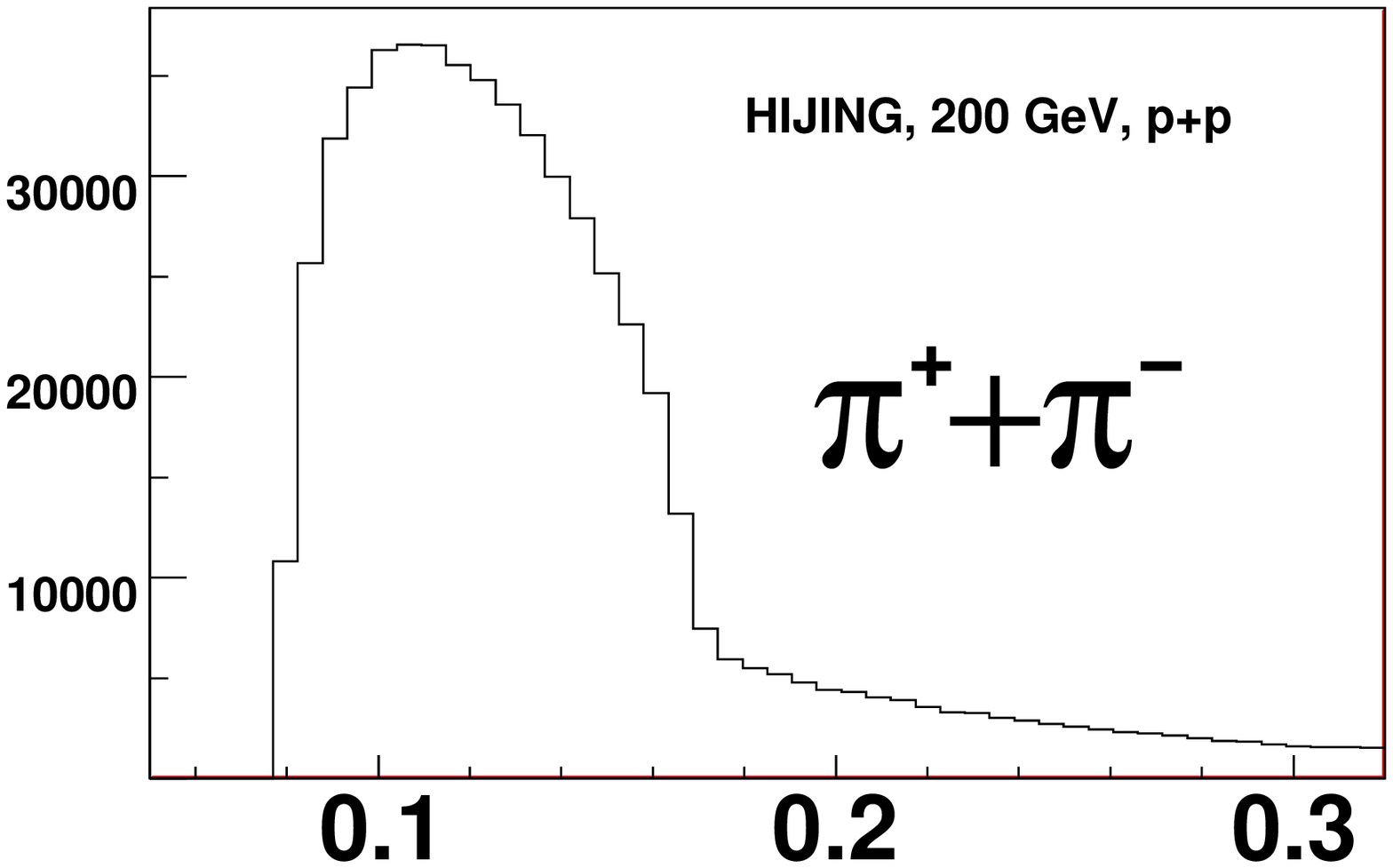}
\includegraphics[width=0.45\linewidth]{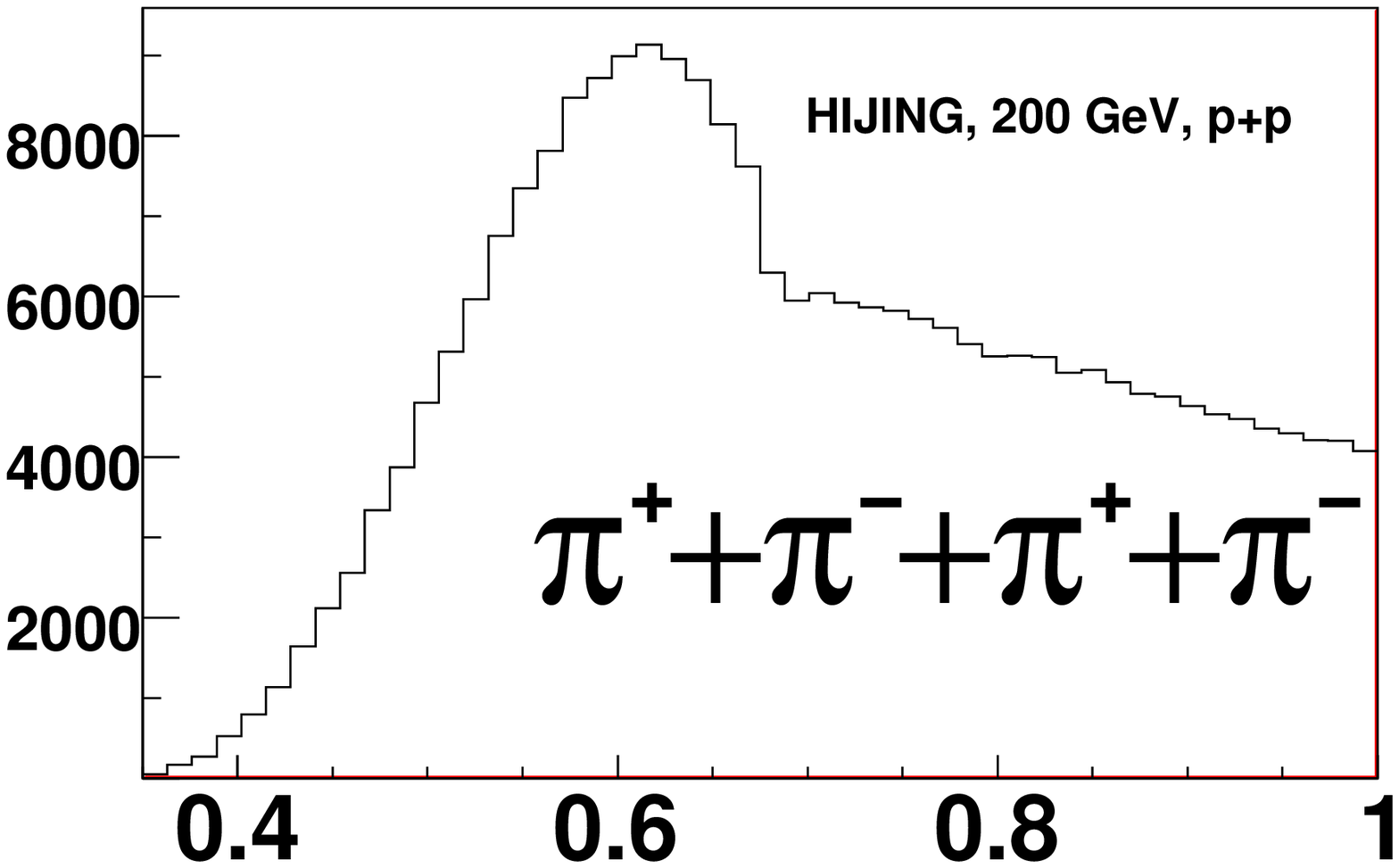}\\
\includegraphics[width=0.45\linewidth]{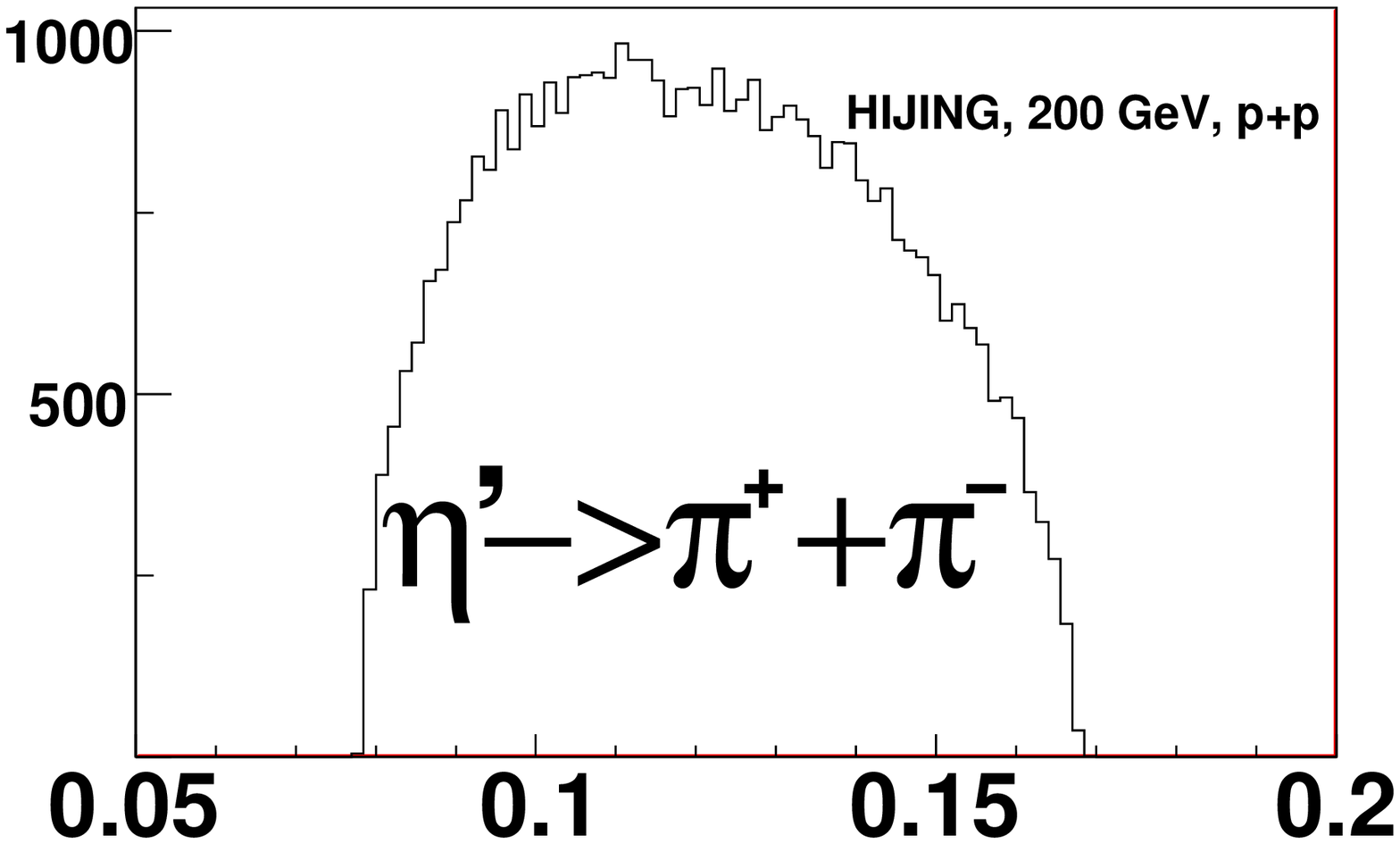}
\includegraphics[width=0.45\linewidth]{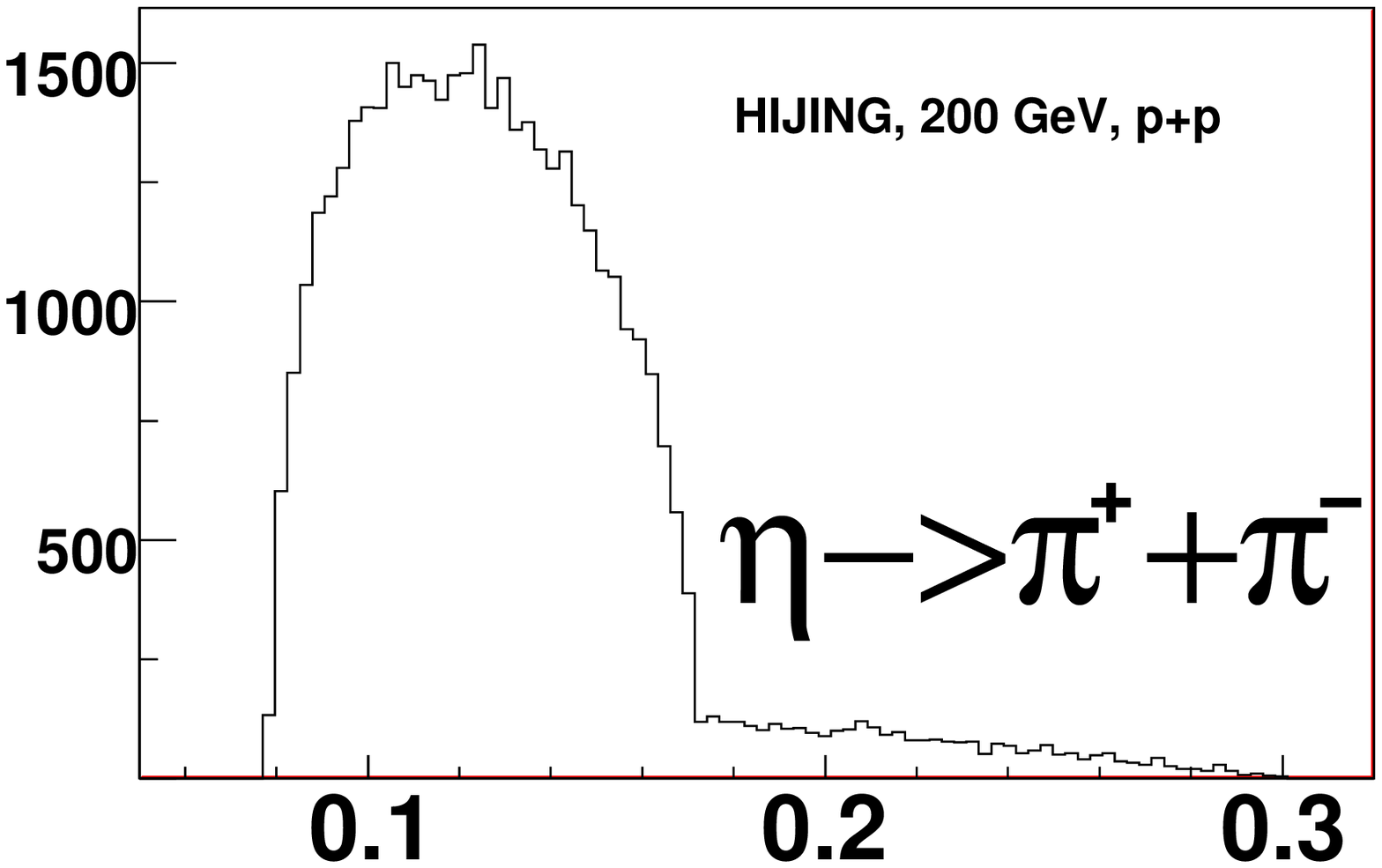}\\
\includegraphics[width=0.45\linewidth]{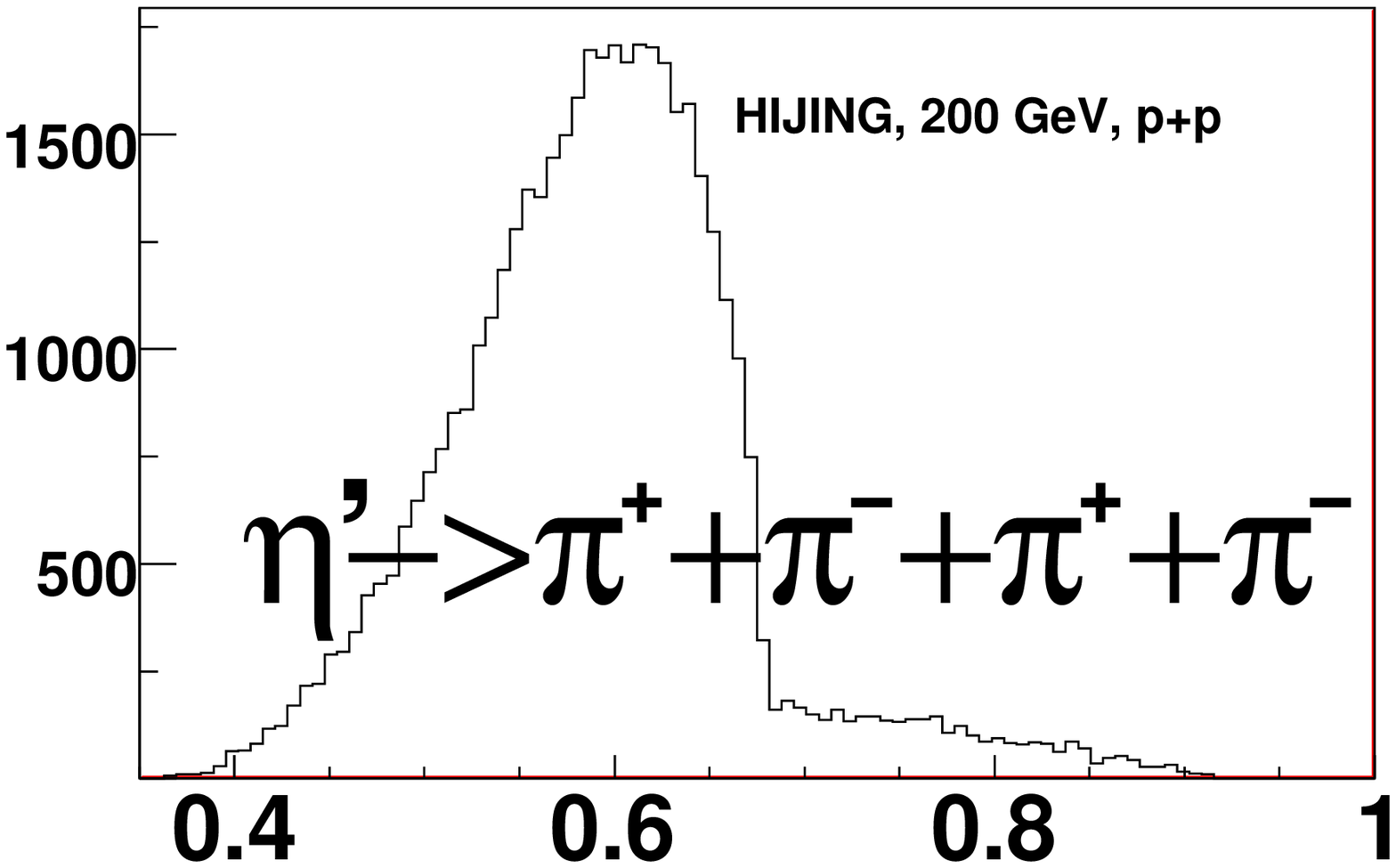}
\end{center}
\caption{Invariant mass distributions for $\sqrt{s}=$200 GeV p+p collisions from Hijing simulations. The first (second) plot shows the
         $m_{inv}$ distribution of all pairs (quadruplets). The last three plots show the $m_{inv}$ distribution for pairs and quadruplets
         coming from $\eta'$ (or from the $\eta$ of the same decay).}
\label{f:minvs}
\end{figure}

Our method is based on the invariant mass of pions from the decay chain of eq.~(\ref{e:decay}). The invariant mass of pion pairs in this decay is:
\begin{align}
m_{inv}^2={\left(E_1+E_2\right)}^2-{\left(p_1+p_2\right)}^2
\end{align}
with $E_1$ and $E_2$ being the energy of the pions, $p_1$ and $p_2$ their three-momentum. Using $E^2=p^2+m^2$ we get
\begin{align}
m_{inv}^2=&m_1^2+m_2^2+2\sqrt{m_1^2+p_1^2}\sqrt{m_2^2+p_2^2}-2p_1p_2\cos{\varphi}\nonumber\\
=&2m_{\pi}^2+2\sqrt{m_{\pi}^2+p_1^2}\sqrt{m_{\pi}^2+p_2^2}-2p_1p_2\cos{\varphi}
\label{e:m2}
\end{align}
where $\varphi$ is the angle between the two pions. If being in the rest system of the $\eta'$, $E_{\eta'}=m_{\eta'}$ holds, and due to momentum conservation
$p_{\eta}=-p_1-p_2$ is also true, so we get
\begin{align}
m_{\eta'}=&\sqrt{m_{\pi}^2+p_1^2}+\sqrt{m_{\pi}^2+p_2^2}+\nonumber\\
&\sqrt{m_{\eta}^2+p_1^2+p_2^2+2p_1p_2\cos{\varphi}}
\label{e:m2etap}
\end{align}
which can be substituted into eq.~(\ref{e:m2}). The $\eta$ can take most of the energy if
$p_1=p_2$ (thus $\varphi=0$), and it has the least energy if $\varphi=\pi$. This yields a lower and an upper bound for
$m_{inv}^2$, and the result for the interval will be $0.078$--$0.168$ GeV$^4/c^2$. Similarly for the second part of the decay chain the same
calculation can be done, and we get $0.078$--$0.166$ GeV$^2/c^4$. This can be checked in
our simulations, and the intervals could been verified. Based on the simulations, we chose the $0.075$--$0.171$ GeV$^2/c^4$ interval for all pairs.
Thus this interval provides a selection method of $\eta'$ decay pions. However, it is not the most effective, since a significant fraction of all other pion pair
is also in this interval. Thus we also checked the invariant mass of all four pions from this decay. It falls in the $0.43$--$0.69$ GeV$^2/c^4$
interval, which, together with the two-pion invariant mass cut, yields an effective method of kinematical selection of $\eta'$ decay pions. See example plots
on fig.~\ref{f:minvs}, from 200 GeV center-of-mass energy p+p collisions, simulated with HIJING 1.411.

\section{Method of tagging $\eta'$ decay pions}

Our method, based on the previous section, is the following. Let us have for example a $\pi^+$ from the experimental sample. Let us then select
any $\pi^-$, $\pi^+$,$\pi^-$ to form a quadruplet. We then check if the invariant mass of the $+$,$-$ pairs is in the above mentioned interval, and
also check if the invariant mass of the quadruplet is in the mentioned interval for quadruplets. If for our starting $\pi^+$ there exist
three other pions (one same sign and two opposite sign) with which all invariant mass criteria are fulfilled, we tag this pion ``found'', as quite
probably it comes from an $\eta'$ decay. A different method is, when we start from a $\pi^+$,$\pi^+$ pair and select an opposite sign pair to
form a quadruplet. Then we check if the quadruplet and the two $+$,$-$ pairs also fulfill the invariant mass criteria, and if yes, we tag this
pair ``found''. This latter method is expected to have one advantage compared to the single particle method: the probability of ``finding'' a
non-$\eta'$ pair than a non-$\eta'$ pion is expected to be smaller. If applied to experimental data, tagged pairs or pions can be removed
from the data thus creating a sample poor in $\eta'$ decay pions.

In simulations, we can determine if the pair of the particle comes from an $\eta'$ decay, so the efficiency of the selection method
can be tested. Whether using the pair or the single particle method, we can form four different groups of them:
\begin{itemize}
\item[a)] Comes from an $\eta'$ and fulfills the $m_{inv}$ criteria
\item[b)] Comes from an $\eta'$ and does not fulfill the $m_{inv}$ criteria
\item[c)] Does not come from an $\eta'$ and fulfills the $m_{inv}$ criteria
\item[d)] Does not come from an $\eta'$ and does not fulfill the $m_{inv}$ criteria
\end{itemize}
Here fulfilling the $m_{inv}$ criteria means that, with our pair or particle, a quadruplet can be formed that fulfills all $m_{inv}$ criteria.
Ideally all $\eta'$ pions go into the first group, while all others fall into the last group. Let us call the number of pions (or pairs in the other method)
in the four groups $N_a$, $N_b$, $N_c$ and $N_d$, respectively. Then the number of all pions (pairs) is $N_a+N_b+N_c+N_d$, while all pions (pairs) from an $\eta'$ are
$N_a+N_b$, and the ratio of the number of all pions (pairs) from an $\eta'$ over the number of all others is $(N_a+N_b)/(N_c+N_d)$ before the filtering.
This ratio is $N_b/N_d$ after applying our method. Both ratios are important, since they are directly connected to the $\lambda$ parameter.
The ``goodness'' of our method is basically the double ratio $(N_a+N_b)/(N_c+N_d)$ over $N_b/N_d$, as this shows whether the cleaned sample
contains less $\eta'$ decay pions.

Note that the kinematic acceptance of our detectors clearly distorts our method. It is possible that not all four pions from an $\eta'$ are detected,
thus the required quadruplet cannot be formed in the sample. In this case not all pions (pairs) will be found (and filtered out), this is what we
call efficiency (and can be calculated as $N_a/(N_a+N_b)$. Also, if the sample is significantly large, there will be a high probability of random
quadruplets to fulfill the $m_{inv}$ criteria. This we call loss, and can be calculated as $N_c/(N_c+N_d)$. This causes our experimental sample
to be smaller, thus we will lose statistics. With proper alignment of the $m_{inv}$ intervals these effects can be minimized. Goal of present
paper is however to investigate the efficiency and loss connected to our the method. A similar method was investigated in
ref.~\cite{Kulka:1990zh} for $e^+ e^-$ collisions. We test the method in p+p and Au+Au collisions, at several center-of-mass energies.

\section{Results}

\begin{figure}
\begin{center}
\includegraphics[width=0.7\linewidth]{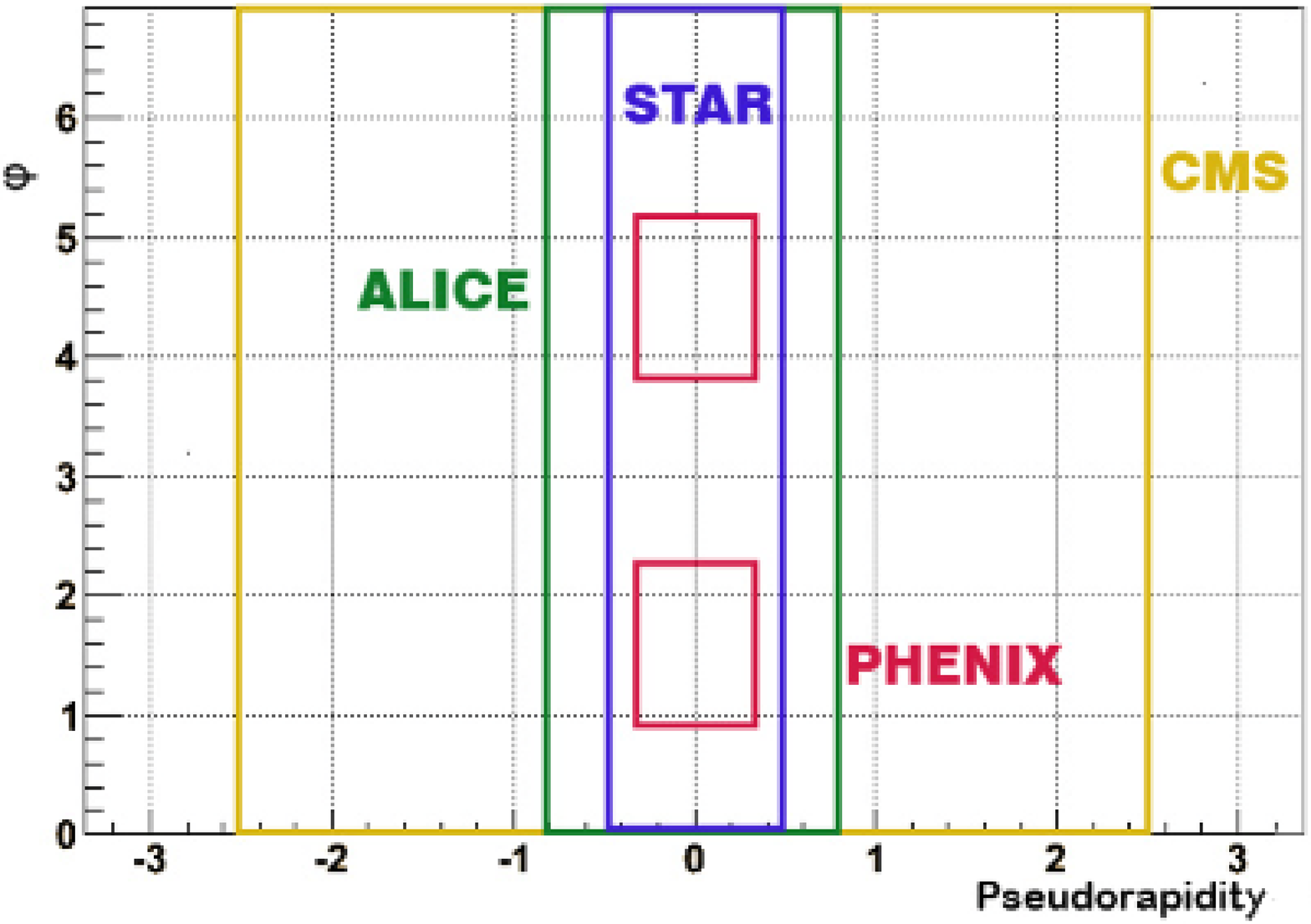}
\end{center}
\caption{Geometrical cuts used in the simulations. PHENIX has the smallest acceptance, thus the efficiency is the worse there.}
\label{f:cuts}
\end{figure}

We used two simulations to test our method: Pythia~\cite{Sjostrand:2007gs} (version 8.135) and Hijing~\cite{Gyulassy:1994ew} (version 1.411).
In the latter, proton-proton and
gold-gold collisions could also be analyzed, while Pythia was used only in case of proton-proton collisions. We also simulated the geometric
acceptance of the detectors. In case of the 200 GeV RHIC energy, we used the geometry of STAR and PHENIX detectors, while in case of
14 TeV energy, we used the geometry of ALICE and CMS detectors. See details on fig.~\ref{f:cuts}.

In the next part, we will give the following three numbers for each type of simulation:
\begin{align}
&\textnormal{Efficiency of finding $\eta'$ pions: }&\frac{N_a}{N_a+N_b}\\
&\textnormal{Loss (found non-$\eta'$ pions): }&\frac{N_c}{N_c+N_d}\\
&\textnormal{Change of $\eta'$ fraction: }&\left.\frac{N_b}{N_d}\middle/\frac{N_a+N_b}{N_c+N_d}\right.
\end{align}
The optimal value for efficiency is 1 (in this case we could cut out all pions coming from $\eta'$ mesons), 0 for loss (in this case
we kept all non-$\eta'$ pions) and 0 for the third, the purification ratio (if it is zero, then after the cut, there are no $\eta'$
decay products at all).

\begin{figure}
\begin{center}
\includegraphics[width=0.95\linewidth]{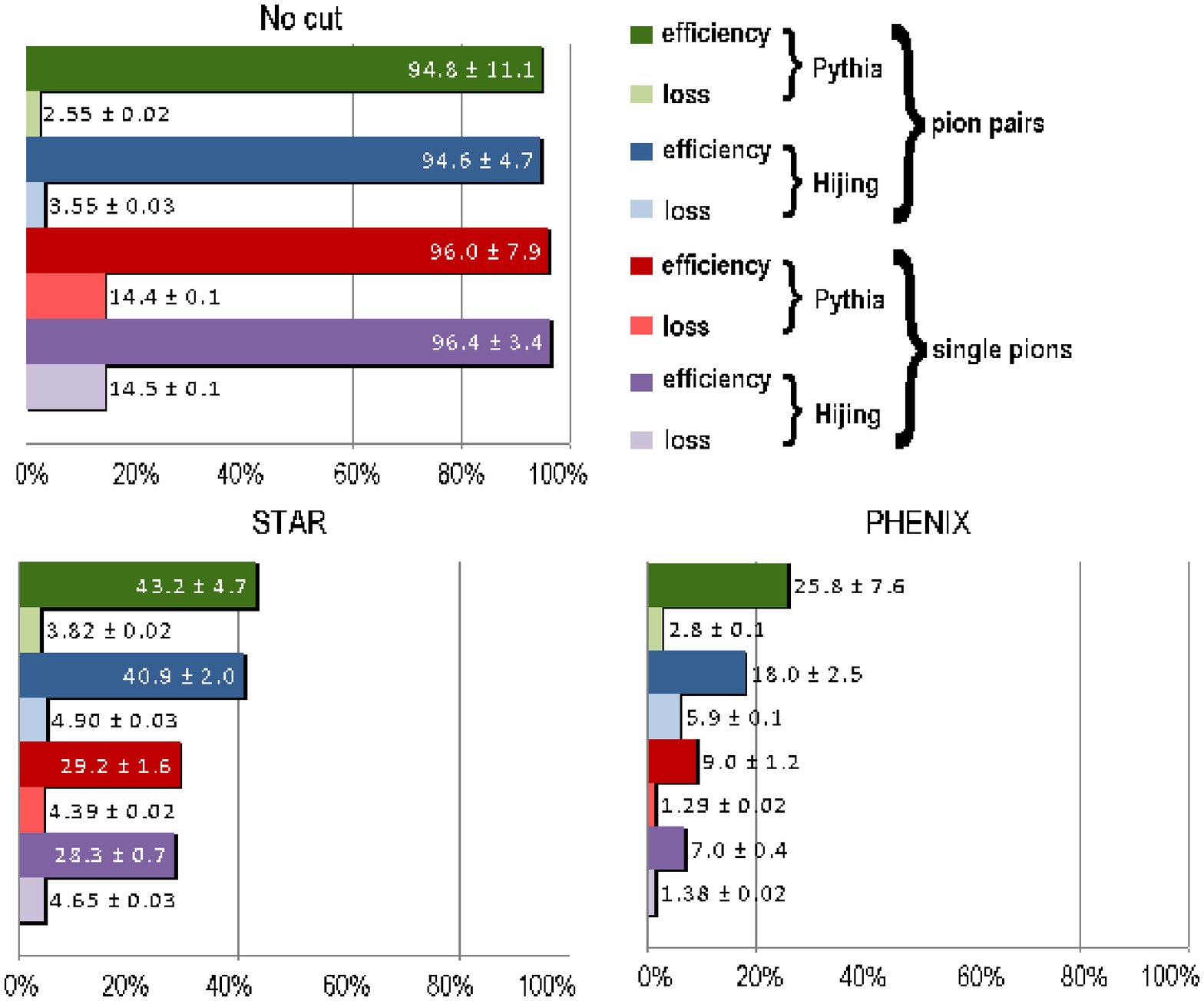}
\includegraphics[width=0.95\linewidth]{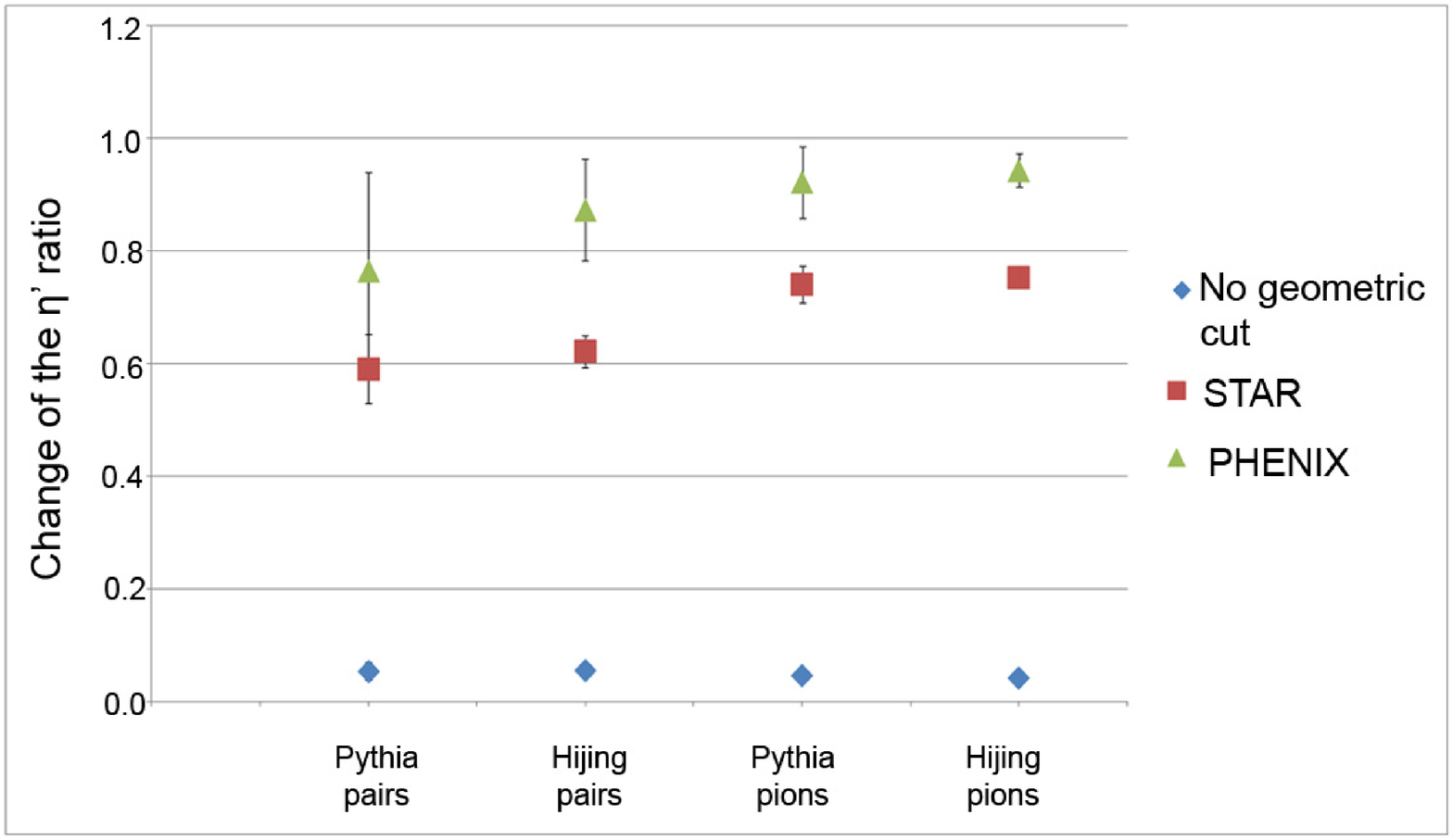}
\end{center}
\caption{Results from $\sqrt{s}=200$ GeV p+p collisions. The method is working in all cases, but efficiency is very low in case of PHENIX cuts,
         for the single particle method.}
\label{f:200pp}
\end{figure}

\begin{figure}
\begin{center}
\includegraphics[width=0.95\linewidth]{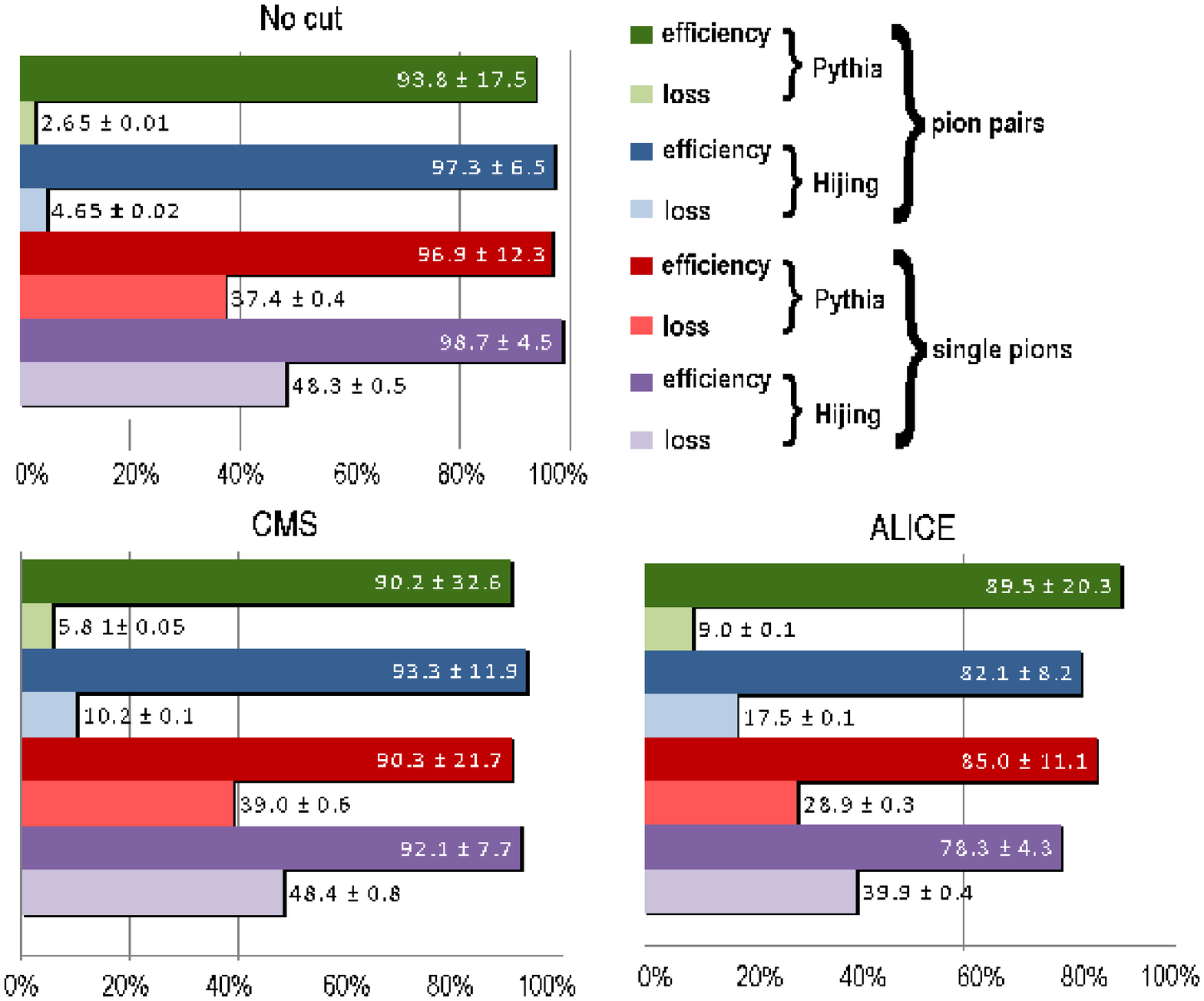}
\includegraphics[width=0.95\linewidth]{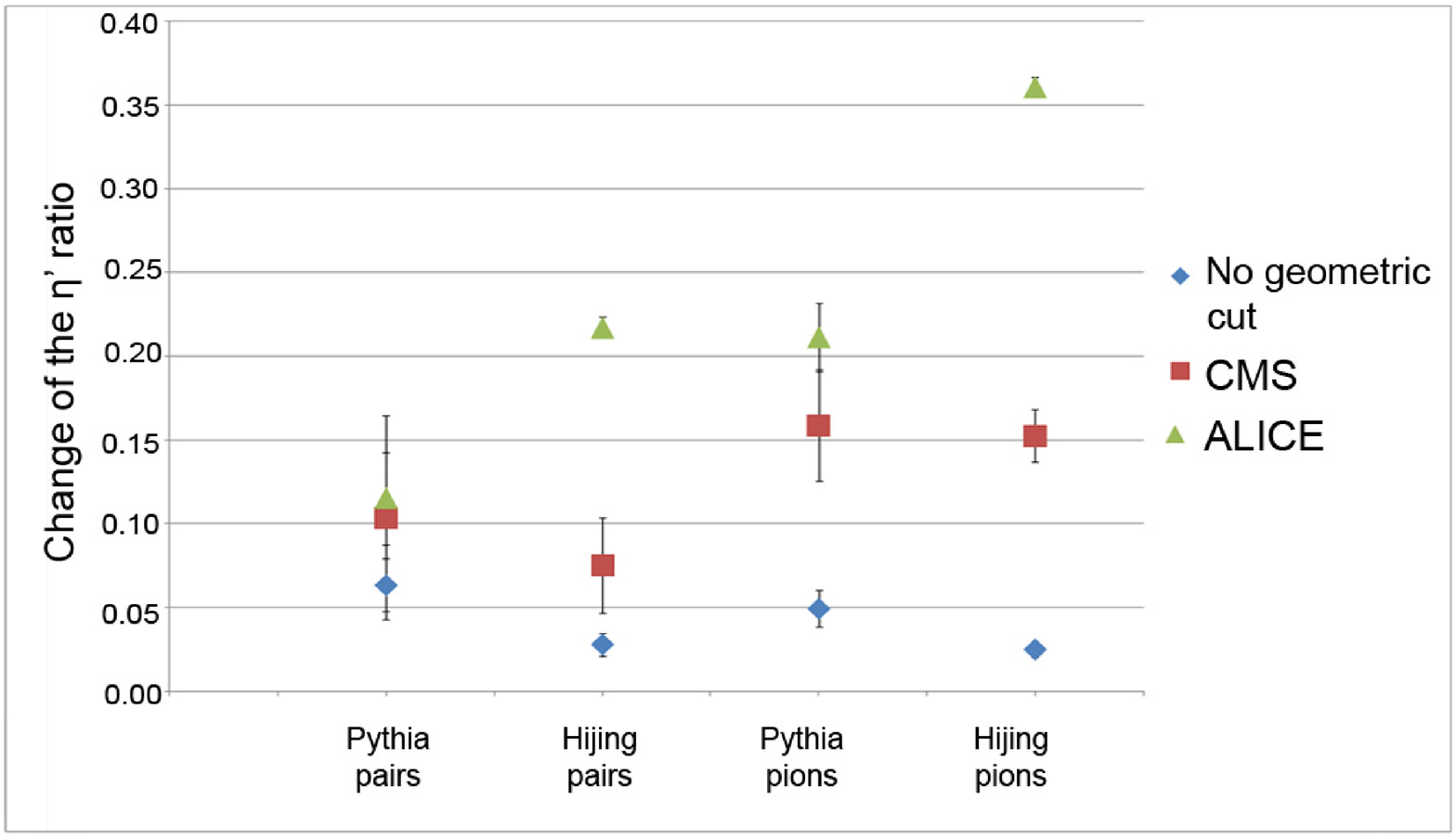}
\end{center}
\caption{Results from $\sqrt{s}=200$ GeV p+p collisions. The method is working in all cases.}
\label{f:14000pp}
\end{figure}

\begin{figure}
\begin{center}
\includegraphics[width=0.95\linewidth]{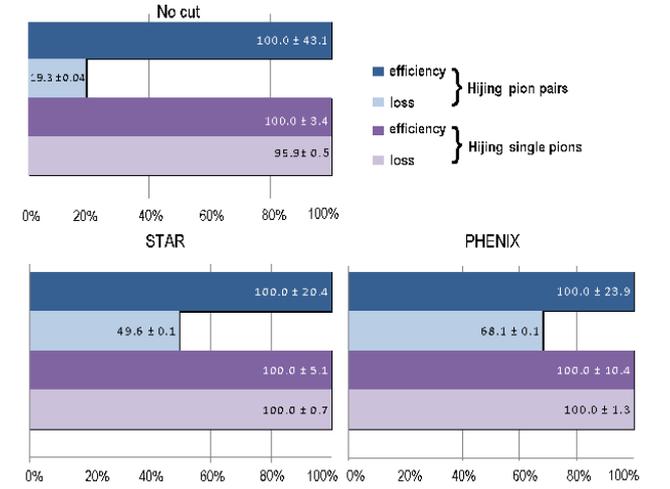}
\end{center}
\caption{Results from $\sqrt{s}=200$ GeV Au+Au collisions. The method is working only in the pair method, as loss is 100\% in the other case.
         We do not give the purification ratio here as due to our 100\% efficiency it always zero (or cannot be calculated if loss is also 100\%).}
\label{f:200auau}
\end{figure}

We generated 1 000 000 p+p $\sqrt{s}=200$ GeV events with both Pythia and Hijing. Geometrical cuts largely influence our method, but it is working in all cases.
However, in case of PHENIX cuts, the efficiency is very low, because a large fraction of pions are not detected, thus we can't ``find''
them to form quadruplets. See fig.~\ref{f:200pp} for details. We also generated 10 000 p+p $\sqrt{s}=14$ TeV events. Here due to the larger average pion number
the efficiency is much larger, and the purification ratio is very good. See details on fig.~\ref{f:14000pp} for details. At both energies, the pair method
is better than the single particle method. We finally generated 100 Au+Au $\sqrt{s_{NN}}=200$ events with Hijing. Here only the pair method was  working, as
essentially all single particles are found, due to the very large statistics. See fig.~\ref{f:200auau} for details. Note that if we ``find'' a pair or particle
that is not coming from
an $\eta'$, then it will be a loss for us, as it decreases our experimental sample. For example if the loss is 50\%, then the sample is reduced by a factor of 2,
so the statistical errors will be increased by a factor of $\sqrt{2}$.

\section{Summary}
We investigated a method of rejecting $\eta'$ decay products in an experimental sample from high energy collisions.
Such a method, if used experimentally, would help to validate signs of partial chiral symmetry restoration.
Our method proposes kinematic cuts in the invariant mass spectrum of pion pairs and quadruplets. The basic
idea is to take a given particle or pair, complement it to a quadruplet with any other random pions,
and check if they fulfill the kinematic criteria. If there is such a complementation for the given pair or
particle, that specific pair or particle can be tagged as coming from an $\eta'$. We find that our method
is working for several systems and several energies. We used to simulations for cross-checking purposes.
The most important system is that of $\sqrt{s_{NN}}=$200 GeV Au+Au collisions, the pair version of
our method is working there. This method can thus be used in an experimental analysis to search
for partial chiral symmetry restoration and the modification of the $\eta'$ mass.

\bibliographystyle{prlstyl}
\bibliography{../../../master}

\begin{thebibliography}{10}

\bibitem{Adcox:2004mh}
K. Adcox {\it et~al.}, Nucl. Phys. {\bf A757},  184  (2005)

\bibitem{Adare:2008fqa}
A. Adare {\it et~al.}, Phys. Rev. Lett. {\bf 104},  132301  (2010)

\bibitem{Kapusta:1995ww}
J.~I. Kapusta, D. Kharzeev, and L.~D. McLerran, Phys. Rev. {\bf D53},  5028
  (1996)

\bibitem{Csorgo:2009pa}
T. Cs\"org\H{o}, R. V\'ertesi, and J. Sziklai, Phys.Rev.Lett. {\bf 105},
  182301  (2010)

\bibitem{Fodor:2009ax}
Z. Fodor and S.~D. Katz,  [arXiv:0908.3341].

\bibitem{Adare:2009qk}
A. Adare {\it et~al.}, Phys. Rev. {\bf C81},  034911  (2010)

\bibitem{Nakamura:2010zzi}
K. Nakamura {\it et~al.}, J. Phys. {\bf G37},  075021  (2010).

\bibitem{Vance:1998wd}
S.~E. Vance, T. Cs\"org\H{o}, and D. Kharzeev, Phys. Rev. Lett. {\bf 81},  2205
   (1998)

\bibitem{Csorgo:1999sj}
T. Cs\"org\H{o}, Heavy Ion Phys. {\bf 15},  1  (2002)

\bibitem{Csanad:2005nr}
M. Csan\'ad, Nucl. Phys. {\bf A774},  611  (2006)

\bibitem{Vertesi:2009ca}
R. V\'ertesi, T. Cs\"org\H{o}, and J. Sziklai, Nucl.Phys. {\bf A830},  631C
  (2009)

\bibitem{Kulka:1990zh}
K. Kulka and B. Lorstad, Nucl. Instrum. Meth. {\bf A295},  443  (1990).

\bibitem{Sjostrand:2007gs}
T. Sjostrand, S. Mrenna, and P.~Z. Skands, Comput.Phys.Commun. {\bf 178},  852
  (2008)

\bibitem{Gyulassy:1994ew}
M. Gyulassy and X.-N. Wang, Comput. Phys. Commun. {\bf 83},  307  (1994)

\end{thebibliography}

\end{document}